\documentclass[conference]{setup/IEEEtran}

\input{setup/preamble.tex}

\begin{document}

\title{Performance Evaluation of Dynamic Metasurface Antennas: Impact of Insertion Losses and Coupling \\
\thanks{This work has been submitted to IEEE for possible publication. Copyright may be transferred without notice, after which this version may no longer be accessible. This work has been supported by Huawei Technologies Sweden AB. 
}}

\author{Pablo Ram\'irez-Espinosa\IEEEauthorrefmark{1}, \IEEEauthorblockN{Robin Jess Williams\IEEEauthorrefmark{1}, Jide Yuan\IEEEauthorrefmark{1} and Elisabeth de Carvalho\IEEEauthorrefmark{1}}
\IEEEauthorblockA{\IEEEauthorrefmark{1}Department of Electronic Systems, Connectivity Section (CNT) Aalborg University, Denmark\\
Email: \IEEEauthorrefmark{1}\{pres, rjw, jyu, edc\}@es.aau.dk
}}

\maketitle

\begin{abstract}

This paper evaluates the performance of multi-user massive multiple-input multiple-output (MIMO) systems in which the base station is equipped with a dynamic metasurface antenna (DMA). Due to the physical implementation of DMAs, conventional models widely-used in MIMO are no longer valid, and electromagnetic phenomena such as mutual coupling, insertion losses and reflections inside the waveguides need to be considered. Hence, starting from a recently proposed electromagnetic model for DMAs, we formulate a zero-forcing optimization problem, yielding an unconstrained objective function with known gradient. The performance is compared with that of full-digital and hybrid massive MIMO, focusing on the impact of insertion losses and mutual coupling. 

\end{abstract}
\begin{IEEEkeywords}
Dynamic metasurface antennas, large intelligent surfaces,  mutual coupling, mutual admittance, wireless. 
\end{IEEEkeywords}

\section{Introduction}\label{sec:introduction}

With the advance and popularity of metamaterials and metasurfaces, \glspl{DMA} are becoming a potential alternative to classical phased-arrays and \gls{MIMO} systems \cite{Smith2017,Sleasman2016, Shlezinger2021, Schlezinger2019}. Formally, a \gls{DMA} is a collection of sub-wavelength radiating elements (e.g., complementary electric resonators \cite{Smith2017}) deployed on top of a guiding structure, usually one-dimensional waveguides connected to a \gls{RF} chain. By introducing some simple semiconductor devices into each element, the so-called reconfigurability is achieved. Stacking several of these one-dimensional structures leads to a large aperture antenna, while still keeping the necessary number of \gls{RF} chains --- and hence, the cost and complexity --- under control.  

Although very promising, the research on beamforming capabilities and transmission rate of \glspl{DMA} is still in an early stage, mainly due to the lack of proper and realistic models for these structures.  Compared with conventional \gls{FD} \gls{mMIMO} architectures, \glspl{DMA} present several particularities that need to be considered; namely the propagation and reflections inside the waveguides, the mutual coupling between radiating element, and the dependence between the \gls{DMA} configuration and the input impedance of the structure (and thus the insertion losses) \cite{Williams2022}. These phenomena translate into two important considerations for system design: \textit{i)} the equivalent channel depends on the specific \gls{DMA} architecture (how the waveguides and elements are deployed) and its momentaneous configuration; and \textit{ii)} the supplied power to the \gls{DMA} varies with its configuration.
 

Despite their relevance, the current literature consider these effects only partially when addressing the beamforming capabilities of \glspl{DMA}. Some basic beamforming techniques are derived in \cite{Smith2017}, although backwards propagation in the waveguides, mutual coupling and insertion losses are ignored. More communications-oriented works are available in \cite{Wang2019, Schlezinger2019}, where the uplink and downlink performance is evaluated in a multiuser \gls{mMIMO} system. However, only the forward propagation in the waveguides and the Lorentzian response of the elements are accounted for. The same happens with \cite{Zhang2021}, where a \gls{DMA}-based multiuser system is compared with full-digital and hybrid solutions. Despite the promising results, the \gls{DMA} model in  \cite{Zhang2021} is still oversimplified, and further studies are necessary to double check their validity in more realistic conditions. 

Motivated by the lack of proper performance analyses for \gls{DMA} systems, in this work we explore their beamforming capabilities when considering all the aforementioned physical phenomena. Specifically, we formulate a zero-forcing problem based on the electromagnetic model in \cite{Williams2022} that leads to an unconstrained non-convex objective function whose gradient is provided, and compare this solution with conventional \gls{FD} and hybrid \gls{mMIMO} systems. In short, we try to answer two important questions: \textit{how close is the performance of \glspl{DMA} to that of \gls{mMIMO} systems?}, and \textit{what phenomena inherent to \glspl{DMA} should be taken into account to render realistic results?}

\textit{Notation:} Vectors and matrices are represented by bold lowercase and uppercase symbols, respectively. $(\cdot)^T$ denotes the matrix transpose, $(\cdot)^H$ is the transpose conjugate, $\|\mat{A}\|_F$ is the Frobenius norm, and $\|\cdot\|_2$ is the $\ell_2$ norm of a vector. Also, $\mat{I}_n$ is the identity matrix of size $n\times n$, $(\mat{A})_{j,k}$ is the $j,k$-th element of $\mat{A}$, and $(\mat{A})_{k,*}$ and $(\mat{A})_{*,k}$ denote the $k$-th row and the $k$-th column of $\mat{A}$ respectively. Finally, $j = \sqrt{-1}$ is the imaginary number, $(\cdot)^*$ indicates complex conjugate, $\mathbb{E}[\cdot]$ is the mathematical expectation, $\circ$ is the Hadamard product, and $\Re\{\cdot\}$ and $\Im\{\cdot\}$ denote real and imaginary part.

\section{System Model}\label{sec:systemModel}
We consider a generic \gls{mMIMO} setup in which one \gls{BS} serves $M$ users simultaneously in a single cell. All the users are modeled as single-antenna devices equipped with a magnetic dipole, as discussed in the sequel, while the \gls{BS} is equipped with either a \gls{FD} array, digital/analog hybrid array or \gls{DMA}. Moreover, we focus on the downlink, and perfect knowledge of the channel is assumed.

For the sake of generality, the three systems are analyzed from a circuital approach, inheriting the electromagnetic model in \cite{Williams2022}. Hence, all the actors in the system  --- namely transmitters, users and antennas --- are represented as ports in the network. The whole network is thus described by 
\begin{align}
    \begin{bmatrix}\vec{v}_\text{t} \\ \vec{v}_\text{s} \\ \vec{v}_\text{r} \end{bmatrix} = \underbrace{\begin{bmatrix}\mat{Y}_\text{tt} &  \mat{Y}_\text{st}^T & \mat{Y}_\text{rt}^T \\ \mat{Y}_\text{st} & \mat{Y}_\text{ss} & \mat{Y}_\text{rs}^T \\ \mat{Y}_\text{rt} & \mat{Y}_\text{rs} & \mat{Y}_\text{rr} \end{bmatrix}}_{\mat{Y}}  \begin{bmatrix}\vec{j}_\text{t} \\ \vec{j}_\text{s} \\ \vec{j}_\text{r} \end{bmatrix},  \label{eq:CircuitalModel_ZF}
\end{align}
where $\vec{v}_k$ for $k\in\{\text{t}, \text{s}, \text{r}\}$ are the complex magnetic voltage vectors at the different ports (measured in amperes), $\vec{j}_k$ are the magnetic current vectors (measured in volts), and the different admittance matrices $\mat{Y}_{k,k'}$ capture the coupling between the corresponding ports. The subindeces $t$, $s$ and $r$ denote, respectively, transmitters (antennas or \gls{RF} chains), \gls{DMA} elements and users --- see \cite{Williams2022} for a more detailed explanation.

From \eqref{eq:CircuitalModel_ZF}, and applying Ohm's law at the different ports, the currents received at the users out of the reactive near-field are expressed in terms of the transmitted currents as \cite[Eq. (5)]{Williams2022}
\begin{align}
    \vec{j}_\text{r} = &\left(\mat{Y}_\text{r}+\mat{Y}_\text{rr}\right)^{-1}\left(\mat{Y}_\text{rs}\left(\mat{Y}_\text{s}+\mat{Y}_\text{ss}\right)^{-1}\mat{Y}_\text{st} - \mat{Y}_\text{rt}\right)\vec{j}_\text{t},
\end{align}
where $\mat{Y}_\text{s}\in\mathbb{C}^{L\times L}$ and $\mat{Y}_\text{r}\in\mathbb{C}^{M\times M}$ are diagonal matrices with elements $(\mat{Y}_\text{s})_{l,l} = Y_{sl}$ and $(\mat{Y}_\text{r})_{m,m} = Y_{rm}$, i.e., the load admittances at the \gls{DMA} elements (if applies) and the users, respectively. Taking into account that $\vec{j}_\text{t}$ is the input to the system --- i.e., the output of the digital baseband precoding --- we can write $\vec{j}_\text{t} = \mat{B}\vec{x}$, where $\mat{B}\in\mathbb{C}^{N\times M}$ is the precoding matrix and $\vec{x}\in\mathbb{C}^{M\times 1}$ is the vector of symbols intended to the $M$ users, which is assumed to meet $\mathbb{E}[\vec{x}\vec{x}^H] = \sigma_x^2\mat{I}_M$. Then, the received complex symbols are expressed as
\begin{equation}
    \vec{y} = \mat{H}_\text{eq}\mat{B}\vec{x} + \vec{n}, \label{eq:SysModelGeneral}
\end{equation}
with $\vec{n}\sim \mathcal{CN}(\vec{0},\sigma_n^2\mat{I}_M)$ being the noise term and
\begin{align}
    \mat{H}_\text{eq} = \widetilde{\mat{Y}}_\text{r} \left(\mat{Y}_\text{rs}\left(\mat{Y}_\text{s}+\mat{Y}_\text{ss}\right)^{-1}\mat{Y}_\text{st} - \mat{Y}_\text{rt}\right). \label{eq:HeqGeneral}
\end{align}
where\footnote{Compared to \cite[Eq. (61)]{Williams2022}, we here introduce a $\sqrt{\frac{\Re{\mat{y}_\text{r}}}{2}}$ term, such that $\|\vec{y}\|_2^2$ equals the received power.} $\widetilde{\mat{Y}}_\text{r} = \sqrt{\frac{\Re{\mat{y}_\text{r}}}{2}}\left(\mat{Y}_\text{r}+\mat{Y}_\text{rr}\right)^{-1}$.

Assuming that the channel --- and thus the precoding --- remains constant for a sufficiently large number of transmitted symbols\footnote{This is equivalent to the widely-used block fading assumption.}, the \gls{SINR} is expressed as
\begin{equation}
    \gamma_m = \frac{|(\mat{H}_\text{eq})_{m,*}(\mat{B})_{*,m}|^2}{\frac{\sigma_n^2}{\sigma_x^2} + \sum_{k=1,k\neq m}^M|(\mat{H}_\text{eq})_{m,*}(\mat{B})_{*,k}|^2}, \label{eq:SINR_General}
\end{equation}
where the transmitted power can be computed as \cite{Pozar2012}
\begin{equation}
    P_t = \frac{1}{2}\mathbb{E}\left[\Re{\vec{j}_\text{t}^H\vec{v}_\text{t}}\right]=  \frac{\sigma_x^2}{2}\Tr{\Re{\mat{B}^H\mat{Y}_\text{p}\mat{B}}}, \label{eq:Pt_General}
\end{equation}
where $\mat{Y}_\text{p}$ is the admittance matrix at the transmitters, i.e., it is defined as $\vec{v}_\text{t} = \mat{Y}_\text{p}\vec{j}_\text{t}$ and naturally depends on the topology under consideration. In the following, we particularize this generic circuital model for \gls{FD} \gls{mMIMO}, hybrid \gls{mMIMO}, and \gls{DMA} systems.

\subsection{FD and hybrid mMIMO}

In this case, the ports corresponding to the \gls{DMA} elements vanish in \eqref{eq:CircuitalModel_ZF}, and each of the transmitters directly represent one antenna attached to a dedicated \gls{RF} chain, leading to the model in \cite{Williams2021ICC}. For the sake of coherence with the \gls{DMA} case, we also model each antenna as a magnetic dipole on a \gls{PEC} plane. Introducing $L=0$ in \eqref{eq:HeqGeneral} yields
\begin{equation}
    \mat{H}_\text{eq}^\text{fd} = - \widetilde{\mat{Y}}_\text{r} \mat{Y}_\text{rt}.  \label{eq:HeqFD}
\end{equation}
where $\mat{Y}_\text{rr}\in\mathbb{C}^{M\times M}$ represents the mutual coupling between users (diagonal matrix if they are spaced enough), and $\mat{Y}_\text{rt}\in\mathbb{C}^{M\times N}$  captures here the wireless propagation channel. Also, when backscattering is neglected, we have in this case that $\mat{Y}_\text{p}\approx \mat{Y}_\text{tt}$. Introducing this result in \eqref{eq:Pt_General} renders the transmitted power for \gls{FD} \gls{mMIMO} systems. Also, the \gls{SINR} is directly given by introducing \eqref{eq:HeqFD} in \eqref{eq:SINR_General}. Since for every system all the antennas are modeled as magnetic dipoles, we have that the elements of $\mat{Y}_\text{rr}$ are given by \cite[Eqs. (44)-(46)]{Williams2022}. On the other hand, in the \gls{FD} system the coupling between antennas in the \gls{BS} is directly given by the free space admittance (taking into account the \gls{PEC} plane), i.e.,
\begin{align}
    (\mat{Y}_\text{tt})_{n,n'} =  \begin{cases}
    i2\omega \epsilon {G}^{(a)}_{\text{e}2,zz}\left(\vec{r}_n, \vec{r}_{n'}\right)  & n \neq n' \\   k\omega \epsilon/{(3\pi)}        & n = n'
    \end{cases}, \label{eq:Ytt_FD}
\end{align}
with $\omega$ the angular frequency, $k$ the wavenumber, $\epsilon$ the electrical permittivity, and  ${G}^{(a)}_{\text{e}2,zz}(\cdot)$ as in \cite[Eq. (39)]{Williams2022}.

Regarding the hybrid architecture, we here assume for simplicity a fully connected topology, which can be regarded as the upper bound in performance for hybrid \gls{mMIMO}. Hence, the only difference is splitting the precoding matrix $\mat{B}$ into the analog and digital matrices, i.e. the beamforming operation $\mat{B} = \mat{Q}\mat{R}$ is carried out in two steps: \textit{i)} the digital precoding $\mat{R}\in\mathbb{C}^{S\times M}$ at the \gls{RF} chains, and \textit{ii)} the analog phase shifting at the antennas $\mat{Q}\in\mathbb{C}^{N\times S}$ where $\abs{(\mat{Q})_{n,m}} = 1 \, \forall \, n,m$ and $S$ is the number of \gls{RF} chains. 

\subsection{DMA system model}

\label{subsec:DMA_system_model}

\begin{figure}[t]
    \centering
    \includegraphics[width = 0.9\columnwidth]{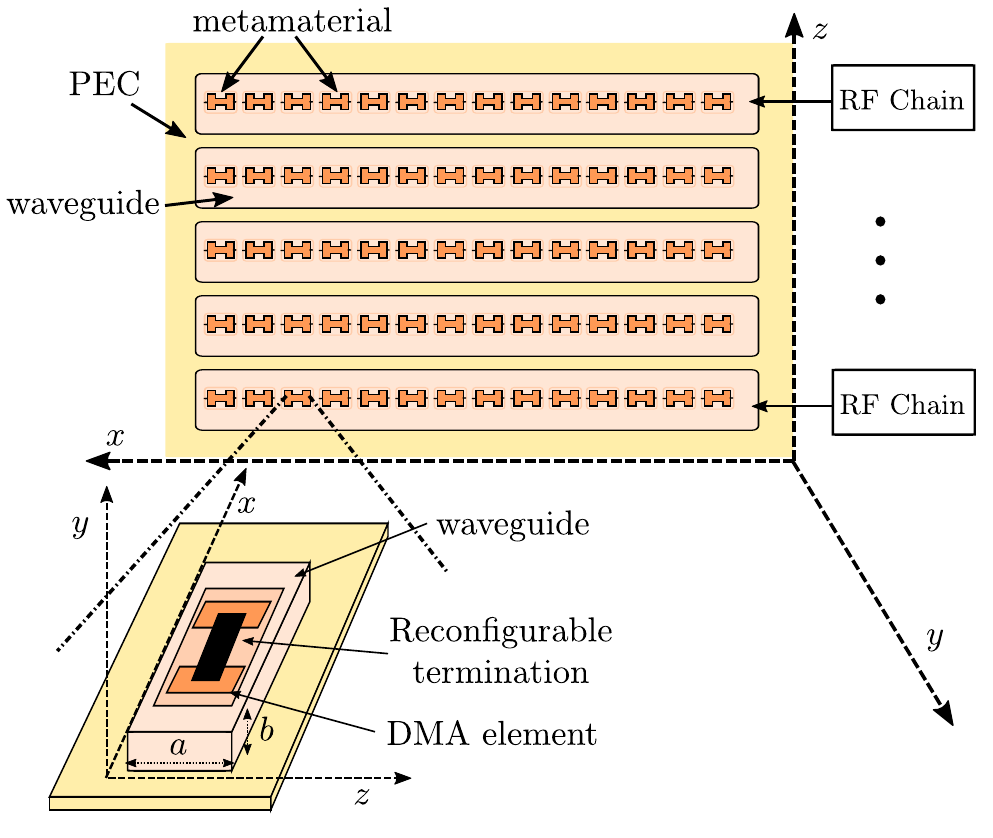}
    \caption{Illustration of a DMA composed by one-dimensional waveguides \cite{Williams2022}.}
    \label{fig:DMA_picture}
\end{figure}

For the \gls{DMA} based system, we consider the model and topology in \cite{Williams2022}, in which the several one-dimensional waveguides are stacked and connected to dedicated \gls{RF} chains, as illustrated in Fig. \ref{fig:DMA_picture}. The transmitter ports in \eqref{eq:CircuitalModel_ZF} no longer represent the antennas at the \gls{BS} but the output of the \gls{RF} chains that feed the waveguides composing the structure. Then, the $L$ radiating elements embedded in these waveguides are characterized by the ports carrying magnetic currents $\vec{j}_\text{s}$. The coupling between the transmitters and the radiating elements are captured in $\mat{Y}_\text{st}\in\mathbb{C}^{L\times N}$, accounting for the propagation and reflections inside the waveguides. The mutual coupling between the antenna elements (occurring both through the air and the waveguides) are described by $\mat{Y}_\text{ss}\in\mathbb{C}^{L\times L}$, and now the wireless channel is given by $\mat{Y}_\text{rs}\in\mathbb{C}^{M\times L}$ (being therefore equivalent to $\mat{Y}_\text{rt}$ in \gls{FD} and hybrid \gls{mMIMO} models). Finally, $\mat{Y}_\text{rt} = \mat{0}$ in this case. Note that the exact expressions for all the admittance matrices are provided in \cite{Williams2022}. Introducing these considerations in \eqref{eq:HeqGeneral}, the equivalent channel is given by
\begin{align}
    \mat{H}_\text{eq}^\text{dma} = \widetilde{\mat{Y}}_\text{r}\left(\mat{Y}_\text{rs}\left(\mat{Y}_\text{s}+\mat{Y}_\text{ss}\right)^{-1}\mat{Y}_\text{st}\right).  \label{eq:HeqDMA}
\end{align}
where, as introduced before, $\mat{Y}_\text{s}\in\mathbb{C}^{L\times L}$ is a diagonal matrix whose elements are the tunable load admittances of the \gls{DMA} elements. The tuning is performed by adjusting the imaginary part of $\mat{Y}_\text{s}$, while the real part $\Re{(\mat{Y}_\text{s})_{l,l}} = R_s$ $\forall$ $l$ represents the parasitic resistance (ideally, $R_s = 0$).  The transmitter admittance matrix $\mat{y}_\text{p}$ is given as 
\begin{align}
    \mat{Y}_\text{p} = \mat{Y}_\text{tt}- \mat{Y}_\text{st}^T\left(\mat{Y}_\text{s}+\mat{Y}_\text{ss}\right)^{-1} \mat{Y}_\text{st}. \label{eq:DMAYp}
\end{align}

Introducing \eqref{eq:HeqDMA} and \eqref{eq:DMAYp} in \eqref{eq:SINR_General} and \eqref{eq:Pt_General} yileds the \gls{SINR} and the transmitted power for the \gls{DMA} system, respectively. However, a key difference with respect to \gls{FD} and hybrid \gls{mMIMO} systems in terms of design is the reflection coefficient at the entry of the waveguides. Since the radiating elements of the \gls{DMA} are connected serially along the feeding waveguide, the insertion losses vary with the tuning of the \gls{DMA}. Therefore, the transmitted (or radiated) power is not enough to characterize the system, and the supplied power $P_s$ is also needed, given by \cite[Eq. (64)]{Williams2022}
\begin{equation}
    P_s =  \frac{\sigma_x^2}{2}\Tr{\Re{\mat{B}^H\p{\mat{I}_N-\mat{\Gamma}^H\mat{\Gamma}}^{-1}\mat{y}_\text{p}\mat{B}}}, \label{eq:Pt_DMA}
\end{equation}
where $\mat{\Gamma}\in\mathbb{C}^{N\times N}$ is a diagonal matrix with the reflection coefficients. Ignoring the effect of the cross-waveguide coupling in the transmitter admittance, the input admittance in the waveguides is written as $\mat{Y}_{\text{in}} = \mat{Y}_\text{p}\circ \mat{I}_N$, rendering from \cite[Eq. (9)]{Williams2022}
\begin{align}
    \mat{\Gamma} = \p{\mat{Y}_{\text{in}} - \mat{i}_N Y_0} \p{\mat{Y}_{\text{in}} + \mat{i}_N Y_0}^{-1},
\end{align}
where $Y_0$ is the characteristic impedance of the source and, therefore
\begin{align}
P_s =&  \frac{\sigma_x^2}{2}\Tr{\Re{\mat{B}^H\mat{y}_\text{q}\mat{B}}}, &
\mat{y}_\text{q} =& \p{\mat{I}_N-\mat{\Gamma}^H\mat{\Gamma}}^{-1}\mat{y}_\text{p}. \label{eq:Y_q}
\end{align}

With this assumption, the reflection coefficient no longer depends on the applied precoding, being convenient for beamforming design, as shown later on. In short, it allows normalizing the transmit power through a direct scaling of the precoding matrix. Note that $P_t \leq P_s$. Since $P_s$ represents the actual power that is consumed by the system, we use it as an optimization constraint.

\section{Zero-Forcing Transmit design}\label{sec:beamforming}

\subsection{FD and hybrid mMIMO}

The zero-forcing solution for a \gls{FD} \gls{mMIMO} system is a well-known result given, e.g., in \cite[Eq. (4.10)]{Bjornson2017}. Denoting by $P_t^\text{max}$ the maximum allowed transmitted power, the constrained precoding matrix is thus written as
\begin{equation}
    \mat{B}_\text{fd} = \frac{\sqrt{P_t^\text{max}}\mat{H}_\text{fd}^\dagger}{ \sqrt{\Tr{\frac{\sigma_x^2}{2}(\mat{H}_\text{fd}^\dagger)^H\Re{\mat{Y}_\text{tt}}\mat{H}_\text{fd}^\dagger}}} \label{eq:B_fd}
\end{equation}
with $\mat{H}_\text{fd}^\dagger = (\mat{H}_\text{eq}^\text{fd})^H\left(\mat{H}_\text{eq}^\text{fd}(\mat{H}_\text{eq}^\text{fd})^H\right)^{-1}$.

For the hybrid architecture, specially for the fully-connected case here considered, several works have proposed beamforming algorithms, most of them focused on approximating the optimal precoding matrix by accounting for the modulus constraint in the phase shift matrix \cite{Yu2016, Ioushua2019, su_optimal_2022}. In our case, we rely on the recently proposed optimal zero-forcing precoder in \cite{su_optimal_2022}, summarized here in \textbf{Algorithm \ref{alg:HybridZF}} for the reader's convenience, where $s$ is the chosen step size for gradient descent and the different matrices involved are computed as
\RestyleAlgo{ruled}
\begin{algorithm}[t]
\caption{Hybrid Zero-Forcing \cite{su_optimal_2022}}
\label{alg:HybridZF}
 \textbf{Input}: $\mathbf{H}^\text{fd}_{\text{eq}}$, $P_t^\text{max}$, $\mat{Y}_\text{tt}$, $s$\;
 \textbf{Initialization}: random $\mat{\Theta}\in \mathbb{R}^{N\times L}$\;
 \While{not converge}{
  Compute RF precoder $\mat{Q}$ and auxiliary matrices $\mat{C}_h$, $\mat{A}_h$, $\mat{U}$, and $\mat{V}$ by \eqref{eq:auxMatrices}\;
  Compute objective gradient $f(\mat{Q})= 2\Im{\left[\mat{U} \p{\mat{A}_h\circ\mat{I}_N - \frac{N}{\frac{P_t^\text{max}}{\sigma^2}+ \Tr{\mat{A}_h} } \mat{I}_N }  \mat{V}\right] \circ \mat{Q}^*}$ \;
  Update $\mat{\Theta} = \mat{\Theta} - s \, f(\mat{Q})$\;
 }
 Compute $\mat{O} =\mat{C} \mat{A}$\;
 Compute diagonal power allocation matrix $\mat{P}$, where $\p{\mat{p}}_{n,n} = \sqrt{\frac{P_t^\text{max} + \Tr{\mat{a}_h}\sigma^2_n}{M \p{\mat{A}_h}_{n,n}}-\sigma^2_n}$ \;
 Compute $\mat{R} = \frac{P_t^{\text{max}} \mat{O}\mat{P}}{\sqrt{\Tr{\frac{\sigma_x^2}{2} \p{\mat{Q}\mat{O}\mat{P}}^H \Re{\mat{Y}_\text{tt}} \mat{Q}\mat{O}\mat{P} }}}$ \;
 \textbf{Outputs}: $\mat{Q}$, $\mat{{R}}$
\end{algorithm}
\begin{align}
\mat{C}_h &= \p{\mat{R}^H \mat{R}}^{-1} \mat{R}^H \p{\mat{H}_\text{eq}^\text{fd}}^H, & \mat{Q} &= \e{i\mat{\Theta}},\notag\\
\mat{A}_h &= \p{\mat{H}_\text{eq}^\text{fd} \mat{R} \mat{C}_h}^{-1}, & \mat{V} &= \p{\mat{C}_h \mat{A}_h}^H,\label{eq:auxMatrices}\\
\mat{U} &= \p{\mat{R}\mat{C}_h - \p{\mat{H}_\text{eq}^\text{fd}}^H} \mat{A}_h. & &\notag
\end{align}

\subsection{DMA zero-forcing beamforming}

Starting from the general communication model in \eqref{eq:SysModelGeneral}, we can cancel the inter-user interference by following the same approach as in \gls{FD} systems in \eqref{eq:B_fd}, setting therefore the baseband precoding matrix to
\begin{equation}
    \mat{B}_\text{dma} = \frac{\sqrt{P_t^\text{max}}\mat{H}_\text{dma}^\dagger}{ \sqrt{\Tr{\frac{\sigma_x^2}{2}(\mat{H}_\text{dma}^\dagger)^H\Re{\mat{Y}_\text{q}}\mat{H}_\text{dma}^\dagger}}} \label{eq:B_dma}
\end{equation}
where $\mat{H}_\text{dma}^\dagger$ is the pseudo-inverse of the \gls{DMA} equivalent channel in \eqref{eq:HeqDMA}, i.e., 
\begin{equation}
    \mat{H}_\text{dma}^\dagger = (\mat{H}_\text{eq}^\text{dma})^H\left(\mat{H}_\text{eq}^\text{dma}(\mat{H}_\text{eq}^\text{dma})^H\right)^{-1}.
\end{equation}
Note that we are considering $\mat{Y}_\text{q}$ in \eqref{eq:Y_q} and, hence, the supplied power instead of the transmitted power, allowing for insertion losses control. Introducing now \eqref{eq:B_dma} in \eqref{eq:SINR_General}, we have that 
\begin{equation}
    \gamma_m  = \gamma = \frac{P_t^\text{max}\sigma_x^2}{\sigma_n^2\Tr{\frac{\sigma_x^2}{2}(\mat{H}_\text{dma}^\dagger)^H\Re{\mat{Y}_\text{q}}\mat{H}_\text{dma}^\dagger}},\quad\forall\,\,m
\end{equation}
where, as expected, the \gls{SNR} is the same for all the users (under the very mild assumption of similar noise power at each user). The beamforming problem is therefore formulated as
\begin{subequations}\label{eq:P1}
\begin{align} 
\mathcal{P}1\;\; \text{(ZF)}:~~ \mathop {\text{maximize}}\limits_{\mat{Y}_\text{s}} \;\;&\mathop {\gamma } \hfill \\
{\text{s.t.}}\;&\Re{ {{{ \p{\mat{Y}_\text{s} }}_{i,i}}} } = R_s \, \forall \, i. \label{eq:constr_ZF} \hfill
\end{align}
\end{subequations}

As defined in Section \ref{subsec:DMA_system_model}, $\mat{Y}_\text{s}$ is a diagonal matrix with the load admittance of each \gls{DMA} element, and whose real part $R_s$  represents the losses. Then, constraint \eqref{eq:constr_ZF} can be easily removed by defining $\mat{Y}_\text{s}^\text{im}\in\mathbb{R}^{L\times L}$ such that  $\mat{Y}_\text{s}^\text{im} = \Im{\mat{Y}_\text{s}}$ and $\widetilde{\mat{Y}}_\text{ss} = R_s\mat{I}_L + \mat{Y}_\text{ss}$. Hence, we can now optimize over the real matrix $\mat{Y}_\text{s}^\text{im}$ while incorporating the losses of the elements in ${\mat{Y}}_\text{ss}$. Moreover, for the sake of notation, we define $\widetilde{\mat{Y}}_\text{rs} = \widetilde{\mat{Y}}_\text{r}\mat{Y}_\text{rs}$, so the \gls{DMA} equivalent channel reads
\begin{equation}
    \mat{H}_\text{eq}^\text{dma} = \widetilde{\mat{Y}}_\text{rs}\left(j\mat{Y}_\text{s}^\text{im}+\widetilde{\mat{Y}}_\text{ss}\right)^{-1}\mat{Y}_\text{st}. \label{eq:H_simp}
\end{equation}

With the above definitions, the optimization problem in \eqref{eq:P1} is finally simplified to
\begin{align}
    &\mathcal{P}2\;\; \text{(ZF)}: \mathop \text{minimize}\limits_{\mat{Y}_\text{s}^\text{im}} \;\;\mathop  f(\mat{Y}_\text{s}^\text{im}) \label{eq:P2}  \\   &\phantom{\mathcal{P}2\;\; \text{(ZF)}:}f(\mat{Y}_\text{s}^\text{im}) = \Tr{\Re{\mat{Y}_\text{q}}\mat{H}_\text{dma}^\dagger(\mat{H}_\text{dma}^\dagger)^H}. \label{eq:CostFunction}
\end{align}

Problem $\mathcal{P}2$ is unconstrained and, in general, non-convex. The difficulty lies on the inverse term $\left(j\mat{Y}_\text{s}^\text{im}+\widetilde{\mat{Y}}_\text{ss}\right)^{-1}$, that appears several times in both $\mat{Y}_\text{q}$ and $\mat{H}_\text{dma}^\dagger$. However, since the problem is unconstrained, approximated solutions can be found by using conventional optimization algorithms available in the literature. In our case, we rely on trust-region methods \cite{Conn2000}, which approximate the objective by a quadratic function in a \textit{trusted region}, i.e., a region where the expected ratio between the expected improvement and the true one is similar. Trust-region optimization is available in standard calculation software such as \textsc{Matlab}, being therefore convenient. Specifically, we use \textit{fminunc} function from \textsc{Matlab}, which implements a simplified trust-region algorithm proposed in \cite{Byrd1988}. The requirement of this algorithm is providing the gradient of the objective function $f(\mat{Y}_\text{s}^\text{im})$, which can be written in closed-form as in \eqref{eq:Gradient} at the top of next page, with the involved auxiliary matrices as
\begin{align}
    \mat{A}^{-1}&=(j\mat{Y}^\text{im}_\text{s}+\widetilde{\mat{Y}}_\text{ss})^{-1}, \label{eq:A}\\
    \mat{C} &= ((\mat{Y}_\text{p}+Y_0)\circ\mat{I}_N)^{-1}, \label{eq:C} \\
    \mat{D} &= (\mat{I}_N-\bm{\Gamma}^H\bm{\Gamma})^{-1}. \label{eq:D}
\end{align}
Due to space constraints, the proof is omitted here, but \eqref{eq:Gradient} is obtained by following standard complex matrix differentiation techniques \cite{Hjrungnes2011}. Note also that, since $\mat{Y}_\text{s}^\text{im}$ is a diagonal matrix, only the diagonal elements of \eqref{eq:Gradient} are necessary.

\begin{figure*}[t]
\begin{align}
    \der{f(\mat{Y}_\text{s}^\text{im})}{\mat{Y}_\text{s}^\text{im}} &= \text{Im}\left\{\left(\mat{A}^{-H}\mat{Y}_\text{st}^*\left[\left((\mat{I}_N-\bm{\Gamma}^H)\bm{\Gamma}\mat{Y}_\text{q}\mat{H}_\text{dma}^\dagger(\mat{H}_\text{dma}^\dagger)^H\mat{D}\mat{C}^H\right)\circ\mat{I}_N\right]\mat{Y}_\text{st}^H\mat{A}^{-H}\right) \right\} \notag \\
    & - \text{Im}\left\{\left(\mat{A}^{-1}\mat{Y}_\text{st}\left[\left(\mat{C}\mat{Y}_\text{q}\mat{H}_\text{dma}^\dagger(\mat{H}_\text{dma}^\dagger)^H\mat{D}\bm{\Gamma}^H(\mat{I}_N-\bm{\Gamma}) + \mat{H}_\text{dma}^\dagger(\mat{H}_\text{dma}^\dagger)^H\mat{D}\right)\circ\mat{I}_N\right]\mat{Y}_\text{st}^T\mat{A}^{-1}\right)\right\} \notag \\
    &- 2\text{Im}\left\{\left(\mat{A}^{-1}\mat{Y}_\text{st}\left[\mat{H}_\text{dma}^\dagger(\mat{H}_\text{dma}^\dagger)^H\Re{\mat{Y}_\text{q}} - (\mat{I}_N-\mat{H}_\text{dma}^\dagger\mat{H}^\text{dma}_\text{eq})\Re{\mat{Y}_\text{q}}\mat{H}_\text{dma}^\dagger(\mat{H}_\text{dma}^\dagger)^H\right]\mat{H}_\text{dma}^\dagger\widetilde{\mat{Y}}_\text{rs}\mat{A}^{-1}\right)\right\}.\label{eq:Gradient}
\end{align}
\hrulefill
\vspace*{4pt}
\end{figure*}

Naturally, the here proposed solution to $\mathcal{P}2$ is not suitable for real-time implementation, and simpler ---albeit suboptimal --- algorithms are necessary. However, the trust-region solver with  \eqref{eq:Gradient} is indeed suitable for performance analysis, providing a reasonable idea on what can be expected from \gls{DMA} based systems. 

\section{Performance evaluation}\label{sec:Simulations}
To compare the performance of the three systems, we run in this section a thorough set of Monte Carlo simulations over correlated Rayleigh channel. Specifically, the channel to the $m$th user is given as $\vec{F}_m\sim \mathcal{CN}(\vec{0},\delta\mat{\Sigma})$ for all $m = 1,\dots,M$.  Resuming the circuital models presented in Section \ref{sec:systemModel}, for the \gls{FD} and hybrid \gls{mMIMO} systems we have $\p{\mat{Y}_\text{rt}}_{m,*} = \vec{F}_m^T$, and for the \gls{DMA} system $\p{\mat{Y}_\text{rs}}_{m,*} = \vec{F}_m^T$. 

The entries of the co-variance matrix $\mat{\Sigma}$ are normalized such that a \gls{FD} system with a single antenna, a conjugate load admittance $Y_{rm} = Y_\text{rr}^*$, and $\sigma_x^2 = 1$ achieves a \gls{SNR} of $\delta$, being therefore the reference system. Hence, using Eqs. \eqref{eq:SysModelGeneral}, \eqref{eq:Pt_General}, and \eqref{eq:HeqFD}, we can write the following relation
\begin{align}
    \Aver[\bigg]{\frac{P_r}{P_t} } = \Aver[\bigg]{\frac{\lvert H_\text{eq}^\text{fd}Bx \rvert^2}{\frac{1}{2}\Re{x^H B^H Y_\text{tt} B x}}} = \frac{9 \delta\sigma_{y}^2 \pi^2}{2 k^2 \omega^2 \epsilon^2},
\end{align}
where $\delta\sigma_y^2$ is the variance of the channel $Y_\text{rt}$ and \eqref{eq:Ytt_FD} and \cite[Eq. (46)]{Williams2022} have been applied. Dividing by the noise variance yields the nominal \gls{SNR} as
\begin{align}
    \overline{\gamma} = \delta\frac{\sigma_y^2}{\sigma_n^2}\frac{9 \pi^2}{2k^2\omega^2\epsilon^2 } \implies \sigma_y^2 = \frac{2k^2\omega^2\epsilon^2\sigma_n^2}{9\pi^2},
\end{align}
such that $\overline{\gamma} = \delta$. Hence, using the covariance matrix in \cite[Eq. (59)]{Williams2022} with $\sigma_\alpha^2 = \frac{2k^2\omega^2\epsilon^2 \sigma_n^2}{9\pi^2} \frac{3}{4\pi}$, and generating the channels using $\vec{F}_m\sim \mathcal{CN}(\vec{0},\delta\mat{\Sigma})$, the nominal \gls{SNR} is given directly by $\delta$. 
Therefore, in the following simulations the channel power is scaled according to that reference and kept constant for all the systems so that $\overline{\gamma} = 13 \text{dB}$. Also, the frequency is set to $ f = \SI{10}{\giga\hertz}$, the width and height of the \gls{DMA} waveguides are, respectively, $a = 0.73\lambda$ and $b = 0.167\lambda$, with $\lambda$ being the wavelength, and the intrinsic admittance at the entrance of the waveguides is $Y_0 = \SI{35.33}{\siemens}$. 
\begin{figure}[t]
    \centering
    \includegraphics[width=0.8\linewidth]{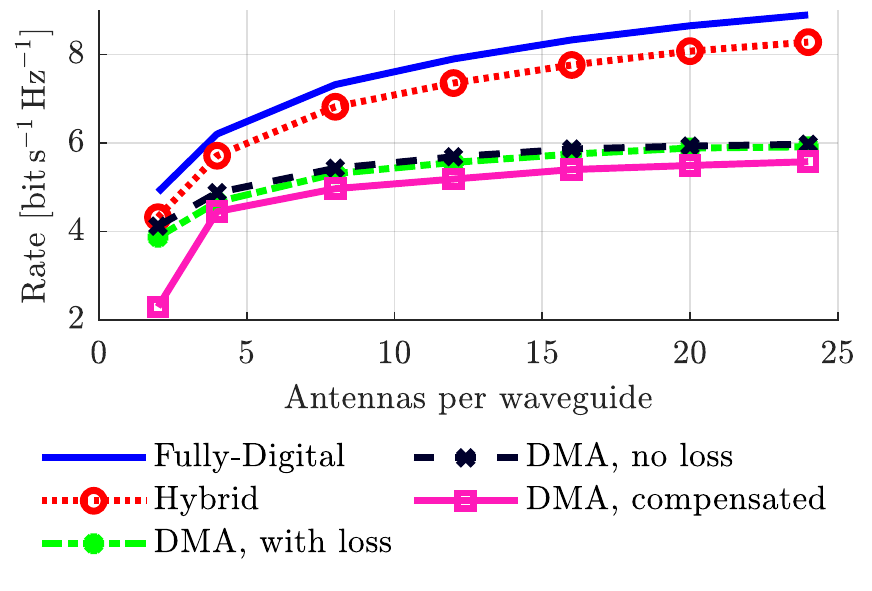}
    \caption{Simulated per user rate for the three presented topologies. There are $M=5$ users being served by $N = 6$ RF chains in the hybrid and {DMA} systems, and the number of antennas connected to each RF chain is increased according to the $x$ axis. For the FD system, the number of RF chains is $N$ times the number of antennas per waveguide. The spacing between the elements is set to $0.5\lambda$, the spacing between waveguides is $\lambda$.}
    \label{fig:Losses}
\end{figure}

The first numerical result is depicted in Fig. \ref{fig:Losses}, where the impact of the insertion losses in the \gls{DMA} system is evaluated. Specifically, for the \gls{DMA} we consider $N=6$ \gls{RF} chains, and start increasing the number of elements per waveguide to see the impact in performance. Three different curves are shown for the \gls{DMA}: \textit{i)} the result obtained for the algorithm in Section \ref{sec:beamforming} (labeled as ``DMA with loss"), \textit{ii)} the performance given by the same algorithm when the reflection coefficient is ignored in the optimization --- we set $\mat{\Gamma} = \mat{0}$ in \eqref{eq:Gradient} --- (labeled as ``DMA, no loss"), and \textit{iii)} the same as before, but the transmitted power is re-scaled so that the supplied power is below $P_t^\text{max}$ (labeled as ``DMA compensated"). 

The performance of the \gls{DMA} system is compared with the hybrid topology with the same number of \gls{RF} chains and antennas, and with the \gls{FD} system when the number of antennas is equal to that in the other systems. Regarding Fig. \ref{fig:Losses}, the first aspect we notice is that, as expected, the \gls{FD} system renders the best performance, but at the cost of a much larger number of \gls{RF} chains. Secondly, we observe that the hybrid solution provides a decent performance while keeping under control the number of \gls{RF} chains; however, note we are using a fully-connected layout, and therefore the required number of phase shifters (and the potential cost) is still high. As for the \gls{DMA} systems, we see that the performance when the insertion losses are considered (green curve) is almost identical to the case without losses (black curve), showing that we can actually control the reflection coefficient in the proposed algorithm. We see, in turn, a drop in performance when the insertion losses are neglected, although the gap seems to reduce as the waveguides are more populated.

Another interesting observation is that the \gls{DMA} system seems to have an upper bound in the achievable rate, in contrast to the \gls{FD} and hybrid \gls{mMIMO} topologies. This is coherent with how a \gls{DMA} works. The power introduced in the waveguides is in part radiated by the first element (according to the termination admittance), and the remaining power keeps propagating until reaching the next element, where the procedure repeats. It is intuitive then that, as we increase the number of elements, less and less power reaches the final elements. One may think that the same happens in \gls{FD} and hybrid systems, but in those topologies, the amplitude and the phase shift induced in the antennas are independent, therefore the power can be equally split between antennas (if this would be necessary). However, in a \gls{DMA}, the amount of power that you radiate by an element and the induced phase shift are related, as explained \cite{Williams2022}. Hence, this leads to the conclusion that may be an optimal number of elements per waveguide in terms of cost and performance ratio.

\begin{figure}[t]
    \centering
    \includegraphics[width=0.8\linewidth]{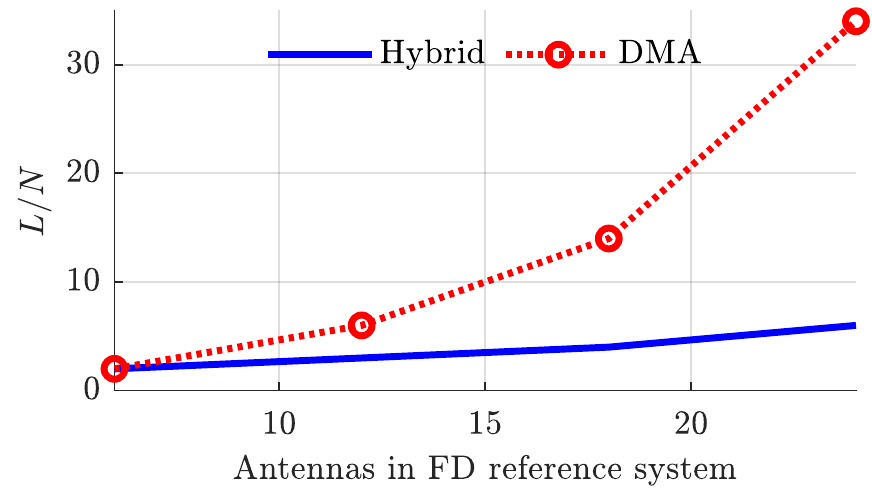}
    \caption{Number of required antennas in hybrid mMIMO and DMA systems to match the performance of a FD array with varying number of antennas.}
    \label{fig:DMA_vs_FD}
\end{figure}

Motivated by the gap in performance between \gls{FD} and hybrid systems and the \gls{DMA}, we explore in Fig. \ref{fig:DMA_vs_FD} how many elements we need to match the performance of a \gls{FD} system. More specifically, we consider a rectangular array in the \gls{FD} case, with $6$ rows and varying number of columns spaced $5\lambda$, increasing thus the number of total antennas. For each number of columns, we explore the number of required antennas in both hybrid and \gls{DMA} systems (denoted by $L$) with $N=6$ \gls{RF} chains. The aperture of the array is kept constant, so to increase the number of elements we reduce the spacing between them. As observed from Fig. \ref{fig:DMA_vs_FD}, while the number of antennas attached to each \gls{RF} chain in hybrid systems seemingly increases linearly, it does exponentially for the \gls{DMA}. This is coherent with the upper bound in performance observed before.  

Finally, we evaluate the impact of mutual coupling in the beamforming solution. To that end, we compare the performance of the proposed zero-forcing algorithm in the following cases: \textit{i)} the \gls{DMA} system as described in Section \ref{sec:systemModel} (labelled as ``DMA"), \textit{ii)} the case where the coupling through the air between the elements is ignored\footnote{The value of $G_{e2,zz}^{(a)}$ when computing \cite[Eq. (37)]{Williams2022} is neglected.} (labelled as ``No air") and \textit{iii)} the mutual coupling is completely ignored, i.e., $\mat{Y}_\text{ss}$ is enforced to be diagonal (labelled as ``No coupling"). 

\begin{figure}[t]
    \centering
    \includegraphics[width=0.8\linewidth]{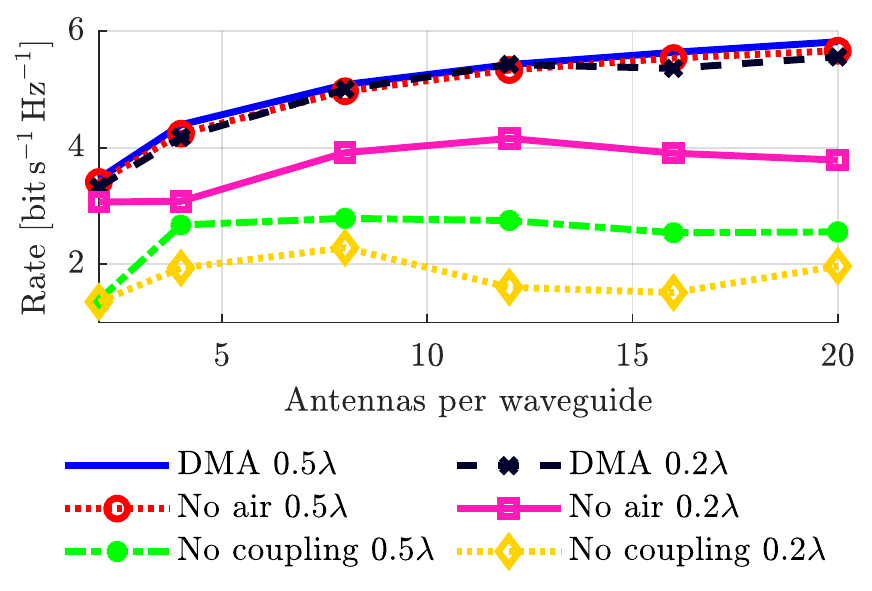}
    \caption{Impact of mutual coupling in simulated per user rate for the DMA system. There are $M=4$ users being served by $N = 4$ RF with different number of elements. The spacing between the elements is set to $0.5\lambda$, the spacing between waveguides is $\lambda$.}
    \label{fig:Coupling}
\end{figure}

The performance of the aforementioned cases is shown in Fig. \ref{fig:Coupling} for two different spacing values between elements: $0.5\lambda$ and $0.2\lambda$. Judging from the results, we clearly see that completely ignoring the coupling leads to a considerably worse performance, while ignoring the coupling through the air is only a reasonable approximation when the spacing between the elements is large. These conclusions are coherent and intuitive, and highlight the necessity of properly modeling the mutual coupling --- both through air and waveguides --- in \gls{DMA} based systems.

\section{Conclusions}\label{sec:Conclusions}
We have provided a complete and electromagnetic-compliant performance evaluation of \gls{DMA} based systems when a zero-forcing beamforming is applied. The results show that, in contrast to \gls{mMIMO} systems, \glspl{DMA} have an upper bound in the achievable rate, and adding more elements per waveguide does not lead to a performance increase once a certain threshold is reached. Also, it has been shown that insertion losses should be considered in the optimization, and that the impact of mutual coupling --- specially that through the waveguides --- is highly relevant. These conclusions are in striking contrast to those observed in other works available in the literature, highlighting the importance of a good modeling when analyzing what can be expected from real systems. 



\bibliographystyle{setup/IEEEtran}
\bibliography{setup/IEEEabrv,setup/references}

\begin{thebibliography}{10}
\providecommand{\url}[1]{#1}
\csname url@samestyle\endcsname
\providecommand{\newblock}{\relax}
\providecommand{\bibinfo}[2]{#2}
\providecommand{\BIBentrySTDinterwordspacing}{\spaceskip=0pt\relax}
\providecommand{\BIBentryALTinterwordstretchfactor}{4}
\providecommand{\BIBentryALTinterwordspacing}{\spaceskip=\fontdimen2\font plus
\BIBentryALTinterwordstretchfactor\fontdimen3\font minus
  \fontdimen4\font\relax}
\providecommand{\BIBforeignlanguage}[2]{{%
\expandafter\ifx\csname l@#1\endcsname\relax
\typeout{** WARNING: IEEEtran.bst: No hyphenation pattern has been}%
\typeout{** loaded for the language `#1'. Using the pattern for}%
\typeout{** the default language instead.}%
\else
\language=\csname l@#1\endcsname
\fi
#2}}
\providecommand{\BIBdecl}{\relax}
\BIBdecl

\bibitem{Smith2017}
D.~R. Smith, O.~Yurduseven, L.~P. Mancera, P.~Bowen, and N.~B. Kundtz,
  ``Analysis of a waveguide-fed metasurface antenna,'' \emph{Phys. Rev.
  Applied}, vol.~8, p. 054048, Nov 2017.

\bibitem{Sleasman2016}
T.~{Sleasman}, M.~F. {Imani}, W.~{Xu}, J.~{Hunt}, T.~{Driscoll}, M.~S.
  {Reynolds}, and D.~R. {Smith}, ``Waveguide-fed tunable metamaterial element
  for dynamic apertures,'' \emph{IEEE Antennas Wireless Propag. Lett.},
  vol.~15, pp. 606--609, 2016.

\bibitem{Shlezinger2021}
N.~Shlezinger, G.~C. Alexandropoulos, M.~F. Imani, Y.~C. Eldar, and D.~R.
  Smith, ``Dynamic metasurface antennas for {6G} extreme massive {MIMO}
  communications,'' \emph{IEEE Wireless Commun.}, vol.~28, no.~2, pp. 106--113,
  2021.

\bibitem{Schlezinger2019}
N.~{Shlezinger}, O.~{Dicker}, Y.~C. {Eldar}, I.~{Yoo}, M.~F. {Imani}, and D.~R.
  {Smith}, ``Dynamic metasurface antennas for uplink massive mimo systems,''
  \emph{IEEE Trans. Commun.}, vol.~67, no.~10, pp. 6829--6843, 2019.

\bibitem{Williams2022}
R.~J. Williams, P.~Ramírez-Espinosa, J.~Yuan, and E.~De~Carvalho,
  ``Electromagnetic based communication model for dynamic metasurface
  antennas,'' \emph{IEEE Trans. Wireless Commun. (Early Access)}, pp. 1--1,
  2022.

\bibitem{Wang2019}
H.~{Wang}, N.~{Shlezinger}, S.~{Jin}, Y.~C. {Eldar}, I.~{Yoo}, M.~F. {Imani},
  and D.~R. {Smith}, ``Dynamic metasurface antennas based downlink massive
  {MIMO} systems,'' in \emph{IEEE 20th Int. Workshop Signal Process. Adv.
  Wireless Commun. (SPAWC)}, 2019, pp. 1--5.

\bibitem{Zhang2021}
H.~Zhang, N.~Shlezinger, F.~Guidi, D.~Dardari, M.~F. Imani, and Y.~C. Eldar,
  ``Beam focusing for near-field multi-user {MIMO} communications,''
  \emph{arXiv preprint arXiv:2105.13087 [eess.SP]}, 2021.

\bibitem{Pozar2012}
D.~M. Pozar, \emph{Microwave engineering}, 4th~ed.\hskip 1em plus 0.5em minus
  0.4em\relax Hoboken, NJ: Wiley, 2012, oCLC: ocn714728044.

\bibitem{Williams2021ICC}
R.~Jess~Williams, P.~Ram{\'i}rez-Espinosa, E.~de~Carvalho, and T.~L. Marzetta,
  ``Multiuser {MIMO} with large intelligent surfaces: Communication model and
  transmit design,'' in \emph{ICC 2021 - IEEE Int. Conf. Commun.}, 2021, pp.
  1--6.

\bibitem{Bjornson2017}
E.~Björnson, J.~Hoydis, and L.~Sanguinetti, \emph{Massive MIMO Networks:
  Spectral, Energy, and Hardware Efficiency}, 2017.

\bibitem{Yu2016}
X.~Yu, J.-C. Shen, J.~Zhang, and K.~B. Letaief, ``Alternating minimization
  algorithms for hybrid precoding in millimeter wave {MIMO} systems,''
  \emph{IEEE J. Sel. Topics Signal Process.}, vol.~10, no.~3, pp. 485--500,
  2016.

\bibitem{Ioushua2019}
S.~S. Ioushua and Y.~C. Eldar, ``A family of hybrid analog–digital
  beamforming methods for massive {MIMO} systems,'' \emph{IEEE Trans. Signal
  Process.}, vol.~67, no.~12, pp. 3243--3257, 2019.

\bibitem{su_optimal_2022}
\BIBentryALTinterwordspacing
X.~Su and Y.~Jiang, ``\BIBforeignlanguage{en}{Optimal {Zero}-{Forcing} {Hybrid}
  {Downlink} {Precoding} for {Sum}-{Rate} {Maximization}},''
  \emph{\BIBforeignlanguage{en}{IEEE Wireless Communications Letters}},
  vol.~11, no.~3, pp. 463--467, Mar. 2022. [Online]. Available:
  \url{https://ieeexplore.ieee.org/document/9634170/}
\BIBentrySTDinterwordspacing

\bibitem{Conn2000}
A.~R. Conn, N.~I.~M. Gould, and P.~L. Toint, \emph{Trust-Region Methods}.\hskip
  1em plus 0.5em minus 0.4em\relax USA: Society for Industrial and Applied
  Mathematics, 2000.

\bibitem{Byrd1988}
R.~H. Byrd, R.~B. Schnabel, and G.~A. Shultz, ``Approximate solution of the
  trust region problem by minimization over two-dimensional subspaces,''
  \emph{Mathematical programming}, vol.~40, no.~1, pp. 247--263, 1988.

\bibitem{Hjrungnes2011}
A.~Hjrungnes, \emph{Complex-Valued Matrix Derivatives: With Applications in
  Signal Processing and Communications}, 1st~ed.\hskip 1em plus 0.5em minus
  0.4em\relax USA: Cambridge University Press, 2011.

\end{thebibliography}

\end{document}